\def\sqrtm1{{\rm i}}
\def\sgn{\,{\rm sign}\,}

\documentstyle[a4,11pt]{article}

\begin{document}

\title{Crystal Distortion and the Two-Channel Kondo Effect}
\author{C. N. Hind and A. O. Gogolin\\Department of 
Mathematics, Imperial College\\ London SW7 2BZ \\ United Kingdom}
\date{9 June 1998}
\maketitle

\begin{abstract}
{We study a simple model of the two-channel Kondo effect in a distorted 
crystal. This model is then used to investigate the interplay of the Kondo and
Jahn-Teller effects, and also the Kondo effect in an impure crystal. We find 
that the Jahn-Teller interaction modifies the characteristic energy scale
of the system below which non-Fermi-liquid properties of the model become 
apparent. The modified energy scale tends to zero as the limit of a purely 
static Jahn-Teller effect is approached. We find also that the non-Fermi-liquid
properties of the quadrupolar Kondo effect are not stable against crystal
distortion caused by impurities.}
\end{abstract}

\section{Introduction}
The general multichannel Kondo model describes a magnetic impurity of spin $S$
in a sea of conduction electrons having spin-$\frac{1}{2}$ and {\em n} 
degenerate orbital channels\cite{one}. Such a model was found by Cox\cite{two} 
 also to apply to a nonmagnetic rare earth impurity such as ${\rm U^{4+}}$ or
${\rm Ce^{3+}}$
with an orbitally degenerate
doublet ground state. In this case, a pseudospin variable associated 
with quadrupolar
deformations of the local orbital plays the part of the impurity spin in the
original multichannel Kondo model, and the real magnetic spin of the 
conduction electrons plays the part of the channel index. Thus we study the
$n=2$ channel Kondo model. This is known as the quadrupolar Kondo effect.

This realisation of the multichannel Kondo model is particularly interesting, 
as the spin symmetry of conduction electrons ensures channel symmetry in the 
model. The channel-symmetric two channel Kondo model has non-Fermi liquid low
temperature behaviour, shown by logarithmic divergence of the impurity 
magnetic susceptibility and specific heat coefficient\cite{one,al,seven}. 

In this paper we study the interplay of the quadrupolar Kondo effect with
local crystal distortion. We consider two possible sources of this distortion:
the Jahn-Teller effect\cite{three} and local distortion caused by non-magnetic
disorder in the crystal. 
For an ion subject to
the quadrupolar Kondo effect, the orbital degeneracy of the ion realises 
conditions for the Jahn-Teller effect: an orbital degeneracy will
split spontaneously on coupling to phonons. This splitting is accompanied by
a distortion of the crystal around the ion. In this case, there will be 
several distortions producing the same splitting of the orbital degeneracy, and
the system may tunnel between these equivalent distortions (the dynamic
Jahn-Teller effect). There has been 
discussion\cite{two,four,five} of this interplay 
of the Jahn-Teller and Kondo effects. It is the intention here to provide a 
simple model of this interplay, and use the technique of Abelian bosonization 
to calculate the low-temperature properties of this model.

Also, other (non-magnetic) impurities in the crystal will cause internal 
strains. The effect of strain on a quadrupolar Kondo ion may be described by a
 term like
$H_1 = gs_z$ in the Hamiltonian, where $g$ is linearly related to components
of the local strain tensor: i.e., it enters the Hamiltonian like a magnetic 
field
term. We investigate the effect of internal strains on the non-Fermi 
liquid behavior of a quadrupolar Kondo impurity, since this behavior is the
main motivation for studying such systems.

\section{The model}
Cox\cite{cx} has shown that many different crystal symmetries are sufficient
to produce a quadrupolar Kondo effect in impurity ions (see also Hirst\cite{H}
for a review of coupling between a local moment and conduction electrons).
 Since we are studying 
the interplay between this effect and local distortions, we choose a simple
crystal symmetry with the hope of producing a tractable model. Thus the model
we consider is one of an impurity ion at a site of tetragonal 
symmetry (lattice lengths $a=b\not=c$). The impurity is taken to be subject to
a quadrupolar Kondo effect. The three-dimensional Kondo Hamiltonian may
be reduced to a one-dimensional Hamiltonian by partial-wave projection (see 
Ref.
\cite{al} Appendix A for a review of this), which in the two channel case
is
\begin{equation}
 H_{\rm K} = \sum_{i=1}^{2} \left\{ H_0[\psi_i] + \frac{I}{2}(s_+J_i^-(0) +
{\rm H.c.}) + Is_zJ_i^z(0) \right\} \ ,
\end{equation}
where 
\begin{eqnarray}
  J_i^-(x) = :\psi_{i\downarrow}^{\dag}(x)\psi_{i\uparrow}(x): \ {\rm and}\  J_i^z(x) = \frac{1}{2}:\psi_{i\uparrow}^{\dag}(x)\psi_{i\uparrow}(x) - \psi_{i\downarrow}^{\dag}(x)\psi_{i\downarrow}(x):
\end{eqnarray}
are the spin currents in the two channels and
\begin{equation}
H_0[\psi_i] = v_{\rm F}\sum_{\sigma = \uparrow,\downarrow}
             \int_{-\infty}^{+\infty}dx 
    :\psi^{\dag}_{i,\sigma}(x)(-{\rm i}\partial_x)\psi_{i,\sigma}(x):
\end{equation}
describes right-moving fermions on ${-\infty}<x<{+\infty}$. This is discussed
in the review by Cox and Zawadowski\cite{cz}.

To this is added a part describing local distortion. The Jahn-Teller
distortion may be described\cite{six} by a pseudospin $\vec{\tau}$. In the 
current case, only the distortion of the four nearest neighbours in the 
$ab$-plane are considered, as only this distortion will couple to the orbitals
 giving rise to the Kondo effect. Thus we may take $\tau_z = \frac{1}{2}$ to 
correspond to the distorted state with an expansion in the x-direction and
contraction in the y-direction. Likewise $\tau_z = -\frac{1}{2}$ corresponds
 to the state with expansion in the y-direction and contraction in the
x-direction.

These distortions are coupled to the pseudospin $\vec{s}$ in $H_{\rm K}$ via a term
\begin{equation}
H_1 = \bar{g}s_z\tau_z
\end{equation}
describing the splitting of the degenerate orbitals by the Jahn-Teller
 distortion. A further term
\begin{equation}
H_2 = \bar{\Delta}_0\tau_x
\end{equation}
is added to describe tunnelling between the states $\tau_z = \pm\frac{1}{2}$.
The total Hamiltonian is then
\begin{equation}\label{ham}
H=H_{\rm K}+H_1+H_2\ .
\end{equation}
Note that to describe the effect of local strain caused by non-Kondo 
impurities, we set $\bar{\Delta}_0 = 0$. $\bar{g}$ is then linearly related to
components of the local strain tensor.

We now follow Emery and Kivelson\cite{seven} and use Abelian bosonization and 
refermionization in the analogue of the Tolouse limit for the one-channel
Kondo model to transform our Hamiltonian (\ref{ham}) to a more tractable form. 
Bosonizing via
\begin{equation}
\psi_{i,\sigma}(x) = \frac{1}{\sqrt{2\pi a_0}}e^{\sqrtm1\sqrt{4\pi}\phi_{i\sigma}(x)}
\end{equation}
and introducing the spin and charge combinations of the fields
\begin{eqnarray}
\phi_{{\rm c}i}(x) & = & \frac{1}{\sqrt{2}}[\phi_{i\uparrow}(x) + \phi_{i\downarrow}(x)]
\nonumber\\
\phi_{i}(x) & = & \frac{1}{\sqrt{2}}[\phi_{i\uparrow}(x) - \phi_{i\downarrow}(x)]
\end{eqnarray}
and the spin and spin-flavour combinations of the spin fields
\begin{eqnarray}
\phi_{{\rm s}}(x) & = & \frac{1}{\sqrt{2}}[\phi_{1}(x) + \phi_{2}(x)]
\nonumber \\
\phi_{{\rm sf}}(x) & = & \frac{1}{\sqrt{2}}[\phi_{1}(x) - \phi_{2}(x)]
\end{eqnarray} 
we obtain
\begin{eqnarray}
H_{\rm K} = H_0[\phi_{\rm s}] + H_0[\phi_{{\rm sf}}] & + & \frac{I}{2\pi a_0}
(s_+e^{\sqrtm1\sqrt{4\pi}\phi_{\rm s}(0)}\cos[\sqrt{4\pi}\phi_{{\rm sf}}(0)]
 + {\rm H.c.}) \nonumber \\
& + & \frac{I}{\sqrt{\pi}}s_z\partial_x\phi_{\rm s}(0)
\end{eqnarray}
where we have dropped the charge fields $\phi_{{\rm c}i}$ as they decouple from the impurity.

Using the canonical transformation
\begin{equation}
U=e^{\sqrtm1\sqrt{4\pi}s_z\phi_{\rm s}(0)}
\end{equation}
\begin{eqnarray}\label{preTol}
H_{\rm K} \longrightarrow U^{\dag}H_{\rm K}U=H_0[\phi_{\rm s}]+H_0
[\phi_{{\rm sf}}] & + &\frac{I}{\pi a_0}s_x\cos[\sqrt{4\pi}\phi_{{\rm sf}}(0)] 
\nonumber \\
& + & \frac{\lambda_+}{\sqrt{\pi}}s_z\partial_x\phi_{\rm s}(0)
\end{eqnarray}
with $\lambda_+=I-2\pi v_f$.

We take $\lambda_+ = 0$ in the following, analagous to taking the Tolouse limit
 in the single-channel Kondo problem. Deviations from $\lambda_+ = 0$ are 
discussed in section 3, and are irrelevant. This has the effect of decoupling 
the $\phi_{\rm s}$ field from the impurity. Defining for notational simplicity 
$\phi(x) = \sqrt{4\pi}\phi_{{\rm sf}}(x)$ and $\bar{\lambda} = 
\frac{I}{\pi a_0}$ we have the bosonised Hamiltonian
\begin{equation}
H=H_0[\phi]+\bar{\lambda} s_x \cos[\phi(0)] + \bar{g}s_z\tau_z+\bar{\Delta}_0\tau_x
\end{equation}

The next step is to refermionise this Hamiltonian: this is similar 
to the refermionization procedure used by Moustakas and Fisher\cite{mf} in 
their study of two-level systems in metals.
 Care must be taken with the
 fermionic representation of the pseudospins $\vec{s}$, $\vec{\tau}$ in order
that the refermionized Hamiltonian be quadratic and that all commutation
relations are satisfied. The standard fermionic
 representation of spins $\vec{s}_1$, $\vec{s}_2$ is
\begin{equation}\label{referm1}
\begin{array}{llllll}
 s_{1+} & = & d_1^{\dag} 
  &,\ s_{1z} & = & d_1^{\dag}d_1 - \frac{1}{2} \\
 s_{2+} & = & d_2^{\dag}e^{i\pi d_1^{\dag}d_1} 
  &,\ s_{2z} & = & d_2^{\dag}d_2 - \frac{1}{2}
\end{array}
\end{equation}
and these components are rotated before identifying $\vec{s}_1\leftrightarrow
\vec{\tau}$ and $\vec{s}_2\leftrightarrow \vec{s}$ so that
\begin{equation}
 \left( \matrix{ \tau_x\cr
                   \tau_y\cr
                   \tau_z} \right)
 = \left( \matrix { s_{1z}\cr
		     s_{1x}\cr
		     s_{1y}}\right),\ 
   \left( \matrix{s_x\cr
		s_y\cr
		s_z} \right)
 = \left( \matrix { s_{2x}\cr
		-s_{2z}\cr
		s_{2y}}\right). 
\end{equation}

 Then refermionizing $\phi$ via
\begin{equation}\label{kleinfs}
e^{\sqrtm1\phi(x)}= \sqrt{2\pi a_0}e^{\sqrtm1\pi (d_1^{\dag}d_1 + d_2^{\dag}d_2)}\psi(x),
\end{equation}
gives the quadratic Hamiltonian
\begin{eqnarray}
H=H_0[\psi] & + & \frac{\bar{\lambda}}{4}\sqrt{2\pi a_0}(d_2^{\dag}-d_2)
[\psi^{\dag}(0) + \psi(0)] \nonumber \\
& + & \frac{\bar{g}}{4}(d_2^{\dag}-d_2)(d_1^{\dag}+d_1) \nonumber \\
& + & \bar{\Delta}_0(d_1^{\dag}d_1 - \frac{1}{2}) 
\end{eqnarray}
where $H_0[\psi]$
describes free right-moving fermions on $-\infty < x < +\infty$.

The factor $\exp [\sqrtm1\pi (d_1^{\dag}d_1 + d_2^{\dag}d_2)]$ introduced in
Equation (\ref{kleinfs}) serves to ensure the commutation relations of the
spins with the electron fields are satisfied. It is precisely the Klein factor
discussed in the review by von Delft and Schoeller\cite{vDS}.

Transforming to Majorana fermions via
\begin{eqnarray}\label{referm2}
 \psi(x) & = & \frac{1}{\sqrt{2}}[\hat{\chi}(x)+\sqrtm1\hat{\xi}(x)] \nonumber\\
 d_1     & = & \frac{1}{\sqrt{2}}(\hat{b} + \sqrtm1\hat{c}) \nonumber \\
 d_2     & = & \frac{1}{\sqrt{2}}(\hat{\eta} + \sqrtm1\hat{a})
\end{eqnarray}
the $\hat{\xi}$-field decouples from the impurity leaving
\begin{equation}\label{Hamiltonian}
H=H_0[\hat{\chi}]+\sqrtm1\lambda \hat{a}\hat{\chi}(0)+\sqrtm1g\hat{a}\hat{b} +
\sqrtm1\Delta_0 \hat{b}\hat{c}
\end{equation}
where $\lambda = - \sqrt{\frac{\pi a_0}{2}}\bar{\lambda}$, $g=-\frac{1}{2}\bar{g}$ 
and $\Delta_0 = \bar{\Delta}_0$.

 Impurity Green's functions, defined as
\begin{equation}
D_{ij}(t)=-\sqrtm1\langle T\zeta_i(t)\zeta_j(0)\rangle
\end{equation}
where $\vec{\zeta} = (\hat{a},\hat{b},\hat{c})$, can be calculated from the
 Hamiltonian (\ref{Hamiltonian}) by e.g. the equation-of-motion method. In 
Matsubara formulation we obtain
\begin{eqnarray}\label{gf}
D_{ij}(\sqrtm1\omega_n) & = & \frac{1}{\omega_{n}^{2} +\Delta_0^{2}}
\frac{1}{\sqrtm1\omega_n \frac{\omega_{n}^{2} +g^2+\Delta_0^{2}}{\omega_{n}^{2} +\Delta_0^{2}} - \lambda^2 G^{(0)}(\sqrtm1\omega_n)} \times \\
&& {\small \left[ 
\begin{array}{ccc}
\omega_n^2 + \Delta_0^2 & 
g\omega_n &
g\Delta_0 \\
-g\omega_n &
-\sqrtm1\omega_n[\sqrtm1\omega_n-\lambda^2 G^{(0)}(\sqrtm1\omega_n)] &
-\sqrtm1\Delta_0[\sqrtm1\omega_n-\lambda^2 G^{(0)}(\sqrtm1\omega_n)] \\
g\Delta_0 &
\sqrtm1\Delta_0[\sqrtm1\omega_n-\lambda^2 G^{(0)}(\sqrtm1\omega_n)] &
\omega_n^2 + g^2 +\sqrtm1\omega_n \lambda^2 G^{(0)}(\sqrtm1\omega_n)
\end{array}
\right] \nonumber}
\end{eqnarray}
where
\begin{equation}
G^{(0)}(\sqrtm1\omega_n) = -\frac{\sqrtm1}{2v_{\rm F}}\sgn\omega_n
\end {equation}
is the free conduction electron Green's function.

The impurity contribution to the free energy $F$ may be calculated by 
evaluating the thermodynamic average of $\hat{a}\hat{\chi}(0)$ using 
(\ref{gf}) and integrating over the coupling constant:
\begin{eqnarray}
\delta F & = & \sqrtm1 \lambda\int_0^1 d\alpha \langle\hat{a}\hat{\chi}(0)
\rangle_\alpha \nonumber \\
& = & -\frac{1}{2}T\sum_{\omega_n}\int_0^1 d\alpha \frac{\Gamma\sgn\omega_n}
{\omega_n\frac{\omega_n^2+g^2+\Delta_0^2}{\omega_n^2+\Delta_0^2} + \alpha
\Gamma\sgn\omega_n} 
\end{eqnarray}
where $\Gamma = \frac{\lambda^2}{2v_{\rm F}}$ 

The sum over $\omega_n$ may be changed to an integral in a standard manner:
\begin{equation}\label{free}
F=F_0 + \int_{-\Omega}^{\Omega}\frac{d\omega}{2\pi} f(\omega) \tan^{-1} \left(
\frac{\Gamma}{\omega \left\{ \frac{-\omega^2 + g^2 +\Delta_0^2}{-\omega^2
+ \Delta_0^2} \right\} } \right)
\end{equation}
where $F_0$ is the free energy of the impurity decoupled from conduction 
electrons.  It is necessary to keep a
momentum cutoff $\Omega$ to obtain a finite expression for $F$, but $\Omega$
may be taken to infinity in calculating quantities obtained by differentiating
$F$.

\section{The Jahn-Teller distortion}

The properties of model (\ref{Hamiltonian}) in the case $\Delta_0 \not= 0$
(modelling Jahn-Teller distortions) are very different from those in the case
$\Delta_0=0$ (modelling deformations caused by local strain fields). This 
section is thus devoted to the case $\Delta_0 \not= 0$.

 From (\ref{free}) the impurity contribution to the entropy may be obtained
\begin{eqnarray}
S & = & -\frac{\partial F}{\partial T} \nonumber \\
  & = & S_0 - \int_{-\infty}^{+\infty}\frac{d\omega}{2\pi} \frac{\omega 
e^{\omega}}{(e^{\omega} + 1)^2} \tan^{-1} \left(
\frac{\Gamma}{T\omega \left\{ \frac{-T^2\omega^2 + g^2 +\Delta_0^2}{-T^2\omega^2
+ \Delta_0^2} \right\} } \right) \nonumber \\
  & = & S_0 - \delta S
\end{eqnarray}
where $S_0$ is the entropy of the impurity decoupled from conduction electrons.
This can easily be calculated as
\begin{equation}
S_0= \ln(2) + \ln(e^{-E_+/T}+e^{-E_-/T})+\frac{1}{T}
\frac {E_+e^{-E_+/T} + E_-e^{-E_-/T}}
 {e^{-E_+/T} + e^{-E_-/T}}
\end{equation}
where $E_\pm=\pm\sqrt{g^2+\Delta_0^2}$, which has a low temperature expansion
\begin{equation}
S_0=\ln(2)+\left(\frac{2E_+}{T}+1\right)e^{-2E_+/T}+O\left(\frac{1}{T}
e^{-4E_+/T}\right).
\end{equation}
 
$\delta S$ may be expanded at low temperature by using
\begin{equation}
 \tan^{-1} \left(
\frac{\Gamma}{T\omega \left\{ \frac{-T^2\omega^2 + g^2 +\Delta_0^2}{-T^2\omega^2
+ \Delta_0^2} \right\} } \right)
= \frac{\pi}{2}\,{\rm sign}\,\omega - \tan^{-1} \left(
\frac{T\omega}{\Gamma}\frac{g^2 + \Delta_0^2}{\Delta_0^2} \right)
+ O\left({(T\omega)}^3\right)
\end{equation}
to obtain
\begin{equation}\label{Result}
S=\frac{1}{2}\ln(2) + \frac{\pi T}{6\Gamma}\frac{g^2 + \Delta_0^2}{\Delta_0^2}
+O(T^2).
\end{equation}

Equation (\ref{gf}) may also be used to examine the local impurity 
susceptibility
\begin{equation}
\chi_l(T)= \int_0^\beta d\tau \langle\langle s_z(\tau)s_z(0)\rangle\rangle,
\end{equation}
i.e. the susceptibility to a strain field acting on the impurity only.
From the refermionization prescription in Equations (\ref{referm1}) and 
(\ref{referm2}), we have
\begin{equation}
s_z = \sqrtm1\sqrt{2}\,\hat{a}\hat{b}\hat{c}
\end{equation}
and thus $\langle\langle s_z(\tau)s_z(0)\rangle\rangle$ must be expanded using
Wick's theorem:
\begin{eqnarray}\label{Wicksthm}
\langle\langle s_z(\tau)s_z(0)\rangle\rangle & = -2 &
[D_{ab}(0)D_{ca}(\tau)D_{bc}(0)
-D_{ab}(0)D_{cb}(\tau)D_{ac}(0) \nonumber \\
&&+D_{ab}(0)D_{cc}(\tau)D_{ab}(0)
-D_{ac}(0)D_{ba}(\tau)D_{bc}(0) \nonumber \\
&&+D_{ac}(0)D_{bb}(\tau)D_{ac}(0)
-D_{ac}(0)D_{bc}(\tau)D_{ab}(0) \nonumber \\
&&+D_{aa}(\tau)D_{bc}(0)D_{bc}(0)
-D_{aa}(\tau)D_{bb}(\tau)D_{cc}(\tau) \nonumber \\
&&+D_{aa}(\tau)D_{bc}(\tau)D_{cb}(\tau)
-D_{ab}(\tau)D_{bc}(0)D_{ac}(0)       \nonumber \\
&&+D_{ab}(\tau)D_{ba}(\tau)D_{cc}(\tau)
-D_{ab}(\tau)D_{bc}(\tau)D_{ca}(\tau) \nonumber \\
&&+D_{ac}(\tau)D_{bc}(0)D_{ab}(0)
-D_{ac}(\tau)D_{ba}(\tau)D_{cb}(\tau) \nonumber \\
&&+D_{ac}(\tau)D_{bb}(\tau)D_{ca}(\tau) ]. 
\end{eqnarray}

Since we are interested in the low-temperature properties of the model, we find
the major contributions to this by finding the discontinuous Green's functions,
 Equation (\ref{gf}). At zero temperature, a
 discontinuous $D_{ij}(\sqrtm1 \omega_n)$ will give a $D_{ij}(\tau)$ behaving
as $1/\tau$ at large $\tau$; whereas a continuous $D_{ij}(\sqrtm1 \omega_n)$
 will give a $D_{ij}(\tau)$ behaving at worst as $1/\tau^2$. We find that
$D_{aa}$, $D_{ac}=D_{ca}$ and $D_{cc}$ are discontinuous in the 
$\omega$-representation. Then defining the constants
\begin{eqnarray}
A_1 &=& D_{ab}(0) \nonumber \\
A_2 &=& D_{ac}(0) \\
A_3 &=& D_{bc}(0) \nonumber
\end{eqnarray}
we find
\begin{equation}
\langle\langle s_z(\tau)s_z(0)\rangle\rangle \simeq -2[ A_3^2 D_{aa}(\tau) +
2A_1A_3 D_{ca}(\tau) + A_1^2 D_{cc}(\tau)],
\end{equation}
these terms arising from terms 1, 3, 7 and 13 of Equation (\ref{Wicksthm}).

We obtain large-$\tau$ properties of this by examining small-$\omega$ 
properties of the $D_{ij}(\sqrtm1\omega_n)$. Since
\begin{eqnarray}
D_{aa}(\sqrtm1\omega_n) &=& \frac{1}{\sqrtm1\Gamma}\sgn\omega +O(\omega)
\nonumber \\
D_{ac}(\sqrtm1\omega_n) &=& \frac{g}{\sqrtm1\Gamma\Delta_0}\sgn\omega 
+O(\omega)\\
D_{cc}(\sqrtm1\omega_n) &=& \frac{g^2}{\sqrtm1\Gamma\Delta_0^2}\sgn\omega 
+O(\omega)
\nonumber
\end{eqnarray}
we have at low temperature and large $\tau$
\begin{eqnarray}\label{asymp}
D_{aa}(\sqrtm1\omega_n) &\sim& -\frac{1}{\pi\Gamma\tau} \nonumber \\
D_{ac}(\sqrtm1\omega_n) &\sim& -\frac{g}{\pi\Gamma\Delta_0\tau} \\
D_{cc}(\sqrtm1\omega_n) &\sim& -\frac{g^2}{\pi\Gamma\Delta_0^2\tau}
\nonumber 
\end{eqnarray}

Thus 
\begin{eqnarray}\label{Result2}
\chi_l(T) &\simeq& \frac{2}{\pi\Gamma}\left( A_3 + \frac{g}{\Delta_0} A_1
\right)^2 \int_{\tau_0}^{1/T-\tau_0} \frac{d\tau}{\tau} \nonumber \\
&\simeq& \frac{2}{\pi\Gamma}\left( A_3 + \frac{g}{\Delta_0} A_1 \right)^2 \ln 
\left( \frac{1}{T\tau_0}\right)
\end{eqnarray}
where $\tau_0$ is a cutoff chosen so that the asymptotics (\ref{asymp}) are
applicable (i.e. this result holds for $T \ll \tau_0$). 

We may also study deviation from the Tolouse limit; i.e. $\lambda_+ \not= 0$
in Equation (\ref{preTol}). The operator we are including is
\begin{equation}
O_+ = \frac{\lambda_+}{\sqrt{\pi}}s_z\partial_x\phi_{\rm s}(0)
\end{equation}
and if we refermionize $\phi_{\rm s}$ and transform to Majorana fermions via
\begin{eqnarray}
\psi_{{\rm s}}(x) &=& \frac{1}{\sqrt{2\pi a_0}}  e^{\sqrtm1 \sqrt{4 \pi} 
              \phi_{{\rm s}} (x)} \nonumber \\
\psi_{{\rm s}}(x) &=& \frac{1}{\sqrt{2}}[\hat{\chi}_{{\rm s}}(x) + \sqrtm1 
                        \hat{\xi}_{{\rm s}}(x)]
\end{eqnarray}
we find the following expression for this operator:
\begin{equation}
O_+(\tau) = -\hat{a}(\tau)\hat{b}(\tau)\hat{c}(\tau)
\hat{\chi}_{{\rm s}}(0,\tau)\hat{\xi}_{{\rm s}}(0,\tau).
\end{equation}
At the Emery-Kivelson line the impurity spin $s_z$ is decoupled from the
$\psi_{{\rm s}}$ electrons so the $O_+ - O_+$ correlation function factorises
into the spin-spin correlation and the density-density correlation function
for free electrons.
\begin{equation}
\langle \langle O_+(\tau)O_+(0) \rangle \rangle = 
\langle \langle s_z(\tau)s_z(0) \rangle \rangle \left[ G^{(0)}(\tau) \right]^2
\end{equation}
Since 
\begin{eqnarray}
G^{(0)}(\tau) &\sim& 1/\tau   \nonumber \\
\langle \langle s_z(\tau)s_z(0) \rangle \rangle &\sim& 1/\tau
\end{eqnarray}
at large $\tau$ and zero temperature, we have
\begin{equation}
\langle \langle O_+(\tau)O_+(0) \rangle \rangle \sim 1/\tau^3
\end{equation}
and thus the operator $O_+$ has scaling dimension 3/2 and is irrelevant.

However, as found by Sengupta and Georges\cite{sg} it is this operator which is
responsible for the behaviour of the impurity specific heat. Comparison with
the calculation in Ref. \cite{nine} gives the result (again for $T \ll \tau_0$)
\begin{equation}\label{Result3}
\frac{\partial C_{{\rm imp}}(T)}{\partial T} \simeq \frac{\lambda_+^2}{2 \pi
 \Gamma v_{{\rm F}}^2} \left( A_3 + \frac{g}{\Delta_0}A_1 \right)^2 \ln \left( 
\frac{1}{T \tau_0} \right)
\end{equation}

Comparing Equation (\ref{Result}) with the  $T\rightarrow 0$ entropy
of the two-channel Kondo model\cite{nine}
\begin{equation}
S(0) = \frac{1}{2} \ln 2
\end{equation}
shows that the Jahn-Teller interaction does not change the residual entropy
of the two-channel Kondo model. This residual entropy is an important and
peculiar feature of non-Fermi-liquid two-channel Kondo models. Since it
persists in the presence of a Jahn-Teller interaction, it is clear that the
non-Fermi-liquid behaviour also persists.

The results\cite{nine} for the two-channel Kondo local impurity susceptibility
\begin{equation}
\chi_l^{{\rm K}}(T) \simeq \frac{1}{\pi \Gamma} \ln \left( \frac{\Gamma}{T} 
\right)
\end{equation}
 and specific heat coefficient
\begin{equation}
\frac{\partial C_{{\rm imp}}^{{\rm K}}(T)}{\partial T} \simeq 
\frac{\lambda_+^2} {8 \pi \Gamma v_{{\rm F}}^2} \ln \left( \frac{\Gamma}{T} 
\right)
\end{equation}
show that the log-divergence of these quantities is governed by the
characteristic energy scale of the system, $\Gamma$. Comparison with Equations
(\ref{Result2}) and (\ref{Result3}) show that this divergence persists but with
a modification of the characteristic energy scale to $\tau_0^{-1}$.  We find
as $\Delta_0 \rightarrow 0$,
\begin{equation}
\tau_0 \sim \frac{g^2 + \Delta_0^2}{\Gamma \Delta_0^2}
\end{equation}
and thus the non-Fermi liquid behavior emerges only at lower and lower
temperatures. This is because the 
{\em dynamic} nature of the Jahn-Teller effect preserves symmetry of the 
system, whereas a purely static Jahn-Teller effect - identical to local
distortion of the crystal - breaks the symmetry of the system and destroys
the non-Fermi liquid behavior. 

Thus the Jahn-Teller effect does not destroy the quadrupolar Kondo effect, but 
modifies the characteristic energy scale of the system.

\section{Crystal distortion by impurities}

Setting $\Delta_0 = 0$ in (\ref{Hamiltonian}) leaves us with a model for a
 quadrupolar Kondo impurity sitting in a strain field caused by other 
impurities in the metal. We wish to obtain properties of this model which
have been averaged over all possible distributions of these impurities. 
Thus the local impurity susceptibility will be given by
\begin{equation}
\langle \chi_l(T) \rangle_g = \int P[g] \chi_l(T,g) dg
\end{equation}
where $P[g]$ is the probability distribution of the local strain field, and 
$\chi_l(T,g)$ the impurity susceptibility of the two-channel Kondo model in a
 field, which may be calculated from (\ref{free}). 

The quantity $P[g]$ is calculated in Refs (\cite{ten}) and (\cite{eleven}). It
is Lorentzian at small $g$ and small impurity concentration, with Gaussian
tails at high $g$. For calculational simplicity, we use 
\begin{equation}
P[g] = \frac{a}{\pi} \frac{1}{g^2 + a^2} ,
\end{equation}
i.e. a Lorentzian distribution for all $g$. This is possible as impurities in
large strain fields have very low susceptibility, so the large-$g$ behaviour
of $P[g]$ is unimportant.

$\chi_l(T,g)$ has the properties that at $g=0$, it is logarithmically divergent
as $T \rightarrow 0$, but this logarithmic divergence is rounded off at finite 
$g$, and $\chi_l(T,g\not=0) \rightarrow$ constant as $T\rightarrow0$. Thus 
we may use 
\begin{equation}
\chi_l(T,g) = \ln (T^2 + g^2)
\end{equation}
to obtain qualitative results on the behavior of $\langle \chi_l(T)\rangle_g$.

These approximations give the result
\begin{equation}
\langle \chi_l(T)\rangle_g \simeq \frac{2\pi}{a} \ln (|a| + T)
\end{equation}
i.e. the impurity susceptibility is finite as $T\rightarrow0$. 
Thus the non-Fermi liquid properties of the quadrupolar Kondo effect are not
stable against crystal distortion caused by impurities.

The effect of impurities on the low-temperature behaviour of the impurity 
susceptibility was discussed by Dobrosavljevi\'c {\em et. al.} \cite{DKK}. 
They considered the effect of local density-of-states fluctuations caused by 
the disorder on the susceptibility. They found that these fluctuations 
{\em enhanced} the non-Fermi-liquid properties of the model.
 
The effect we considered above suppresses the non-Fermi-liquid behaviour, and 
is thus in competition with that of Dobrosavljevi\'c {\em et. al.}

\section{Acknowledgments}
The authors would like to thank Y. Chen, M. Fabrizio and A. C. Hewson for 
useful discussions. We acknowledge financial support by the EPSRC.

\end{document}